\documentstyle[prd,eqsecnum,aps]{revtex}

\begin{document}

\draft

\title{Topological Reverberations in Flat Space-times}  

\author{G.I. Gomero\cite{internet1}, M.J. Rebou\c{c}as\cite{internet2}, 
        A.F.F. Teixeira\cite{internet3}, and A. Bernui\cite{internet4} }

\address{Centro Brasileiro de Pesquisas F\'\i sicas \\
	 Departamento de Relatividade e Part\'\i culas \\
	 Rua Dr.\ Xavier Sigaud 150 \\
	 22290-180 Rio de Janeiro -- RJ, Brazil}

\date{\today}

\maketitle

\begin{abstract}
We study the role played by multiply-connectedness in the
time evolution of the energy $E(t)$ of a radiating 
system that lies in static f\/lat space-time  manifolds
${\cal M}_4$ whose $t=const\,$ spacelike sections 
${\cal M\/}_3$ are compact in at least one spatial 
direction. 
The radiation reaction equation of the radiating 
source is derived for the case where ${\cal M}_3$ 
has any non-trivial f\/lat topology, and 
an exact solution is obtained.
We show that the behavior of the radiating energy $E(t)$
changes remarkably from exponential damping, when the 
system lies in ${\cal R}^3$, to a reverberation 
pattern (with discontinuities in the derivative $\dot{E}(t)$ 
and a set of relative minima and maxima) followed by a growth 
of $E(t)$, when ${\cal M\/}_3$ is endowed with any one  
of the 17 multiply-connected f\/lat topologies.
It emerges from this result that the compactness in at 
least one spatial direction of Minkowski space-time is 
suf\/f\/icient to induce this type of topological 
reverberation, making clear that topological fragilities
can arise not only in the usual cosmological modelling, but
also in ordinary f\/lat space-time manifolds.
An explicit solution of the radiation reaction equation
for the case where ${\cal M}_3 = {\cal R}^2 \times 
{\cal S}^1$ is discussed in details, and  graphs which 
reveal how the energy varies with the time are presented 
and analyzed.
\end{abstract}
%\vspace{2mm}
%\pacs{{\bf PACS numbers:} 04.20.Cv, 98.80.Hw, 04.20.Gz, 
%                          98.80.-k, 04.20.Jb, 02.40.-k} 

%%%%%%%%
\section{Introduction}       
\label{intro}
\setcounter{equation}{0}

In general relativity, as well as in any metrical theory
of gravitation of some generality and scope, a common 
approach to cosmological modelling commences with a 
four-dimensional space-time manifold (representing the 
physical events) endowed with a Lorentzian metric
(necessary to ensure the local validity of the well 
established special relativity theory). 
Representing the physical phenomena in this 
manifold we have f\/ields satisfying appropriate local 
dif\/ferential equations (the physical laws). Finally, f\/ields 
and geometry are coupled according to the corresponding 
gravitational theory one is dealing with. The space-time 
geometries arising as solutions of the gravitational f\/ield 
equations constrain to some extent the 
dynamical behavior of the physical f\/ields.  
This eminently metrical approach to model the physical world 
has led a number of physicists to implicitly (or explicitly) 
restrict their studies to purely geometric features of space-time, 
either by ignoring the role of topology or by considering 
just a limited set of topological alternatives for the 
space-time manifold. 

If, on the one hand, the topological properties of a manifold 
antecede and are more fundamental than the metrical features 
and the dif\/ferentiable structure on which tensor analysis is 
based upon, on the other hand, it is well known that geometry 
constrains but it does not dictate the topology of a space-time.
It is therefore important to determine whether and (or) to what 
extent physical results concerning a space-time geometry depend 
on or are somehow constrained by the topology of the underlying
manifold.

In a recent work~\cite{BernuiGomeroReboucasTeixeira98} we 
studied the role played by topology
in the time evolution of the energy $E(t)$ of a 
radiating system in static f\/lat FRW space-time manifolds
${\cal M}_4$ whose  $t=const\,$ spacelike sections 
${\cal M}_3$ are endowed with dif\/ferent topologies, 
namely the simply-connected Euclidean manifold
${\cal R}^3$, and six topologically non-equivalent f\/lat 
multiply-connected orientable {\em compact\/} 3-manifolds%  
~\cite{Wolf67}~--~\cite{Gomero97}. 
Clearly these f\/lat space-time manifolds ${\cal M}_4$ are  
orientable and decomposable into 
${\cal M}_4 = {\cal R} \times {\cal M}_3\,$.
The radiating system we have examined, which has been used to 
study a wide class of oscillating phenomena%
~\cite{SchwalbThirring64}~--~\cite{Bernui94}, 
is formed by a pointlike harmonic oscillator (energy source) 
coupled with a massless scalar f\/ield
(scalar radiation waves propagating at speed of light).
Through a {\em numerical\/} integration of the time evolution 
equation for the harmonic oscillator (with the topological 
constraints suitably considered), we have shown that there
is an exponential damping of the energy $E(t)$ of the harmonic
oscillator when the spacelike section ${\cal M}_3$ is the 
Euclidean space ${\cal R}^3$, whereas when ${\cal M}_3$ 
is endowed with one of the six compact and orientable f\/lat 
topologies
the energy $E(t)$ exhibits a loosely speaking  
reverberation pattern, with discontinuities in the derivative 
of $E(t)$ and relative minima and maxima 
(both the discontinuities and the extrema are due to
 the travelling waves which ``ref\/lect'' from the 
``topological walls''), followed by a growth of the energy 
with the time. 
We have also shown that, for these six compact cases, the energy 
$E(t)$ diverges exponentially  when $t\, \rightarrow \infty\,$, 
in striking contrast with the damping of $E(t)$ in the 
${\cal R}^3$ case.
This unexpected divergent behavior of the energy for 
f\/lat space-times with compact spacelike
sections illustrates that ({\em totally\/}) compact topologies 
may give rise to rather important dynamic changes in the
behavior of a physical system. This type of 
sensitivity has been referred to as topological fragility 
and can occur without violation of any local physical 
law~\cite{ReboucasTavakolTeixeira98}. 

\begin{sloppypar} 
In the present work we extend these investigations by 
performing a rigorous {\em non-numerical\/} study of the 
behavior of the same physical system 
under a less restrictive topological setting, namely we shall 
consider static f\/lat space-time  manifolds ${\cal M}_4$ whose 
$t=const\,$ spacelike sections ${\cal M\/}_3$
are compact in at least one spatial direction.
Moreover, no assumption will be made as to whether or not 
these 3-manifolds  are orientable.
This amounts to saying that the results of the present 
work hold for all f\/lat space-time manifolds
whose $t=const\,$ spacelike sections 
are endowed with one of the 17 f\/lat topologies
discussed by Wolf~\cite{Wolf67}, Ellis~\cite{Ellis71}, and
others~\cite{LachiezeReyLuminet95} 
(see also Gomero~\cite{Gomero97}).
\end{sloppypar}

It should be stressed from the outset that in spite of 
being the same physical system discussed in%
~\cite{BernuiGomeroReboucasTeixeira98}, the present 
work generalizes the results of that article in several
respects. Firstly, the underlying topological 
setting is now much more general in that here we do not assume 
that the spacelike 3-manifolds ${\cal M}_3$ are compact 
and orientable. 
Secondly, contrarily to the {\em numerical\/} integration 
performed in~\cite{BernuiGomeroReboucasTeixeira98}, in the 
present paper we have obtained a closed {\em exact\/} solution 
of the evolution equation for the harmonic oscillator
in the above-mentioned general topological setting.
Thirdly, the case study we have examined in section~\ref{case}, where
we obtain the explicit solution for the radiation 
reaction equation and discuss in details the graphs 
which reveal the time evolutions of $E(t)$ and 
$\dot{E}(t)$, is completely new.
Finally here we have also performed a lengthy discussion on 
the conservation and on the balance of the total energy 
of our physical system, making apparent how (here as well 
as in ref.~\cite{BernuiGomeroReboucasTeixeira98}) the total 
energy of the whole system is conserved for any time $t>0$. 

The plan of this article is as follows.
We derive (section~\ref{RadReacEq}) the time evolution equation 
for the harmonic oscillator (also referred to as radiation reaction 
equation) when ${\cal M}_3\,$ is endowed with a non-trivial 
f\/lat topology, i.e., when the  $t=const$ spacelike sections 
are any multiply-connected (compact in at least one direction)
f\/lat manifold ${\cal M}_3$.
It turns out that the evolution equation for
this case is formally the same as that obtained by 
Bernui {\em et al.\/}~\cite{BernuiGomeroReboucasTeixeira98} 
for the case where ${\cal M}_3$ is any (totally) compact 
and orientable 3-manifold. However, they dif\/fer in 
the non-homogeneous term that accounts for the
dif\/ference in the topology ---~distinct
degrees of connectedness of the spacelike sections
give rise to dif\/ferent non-homogeneous terms.
We also emphasize in section~\ref{RadReacEq} that since our
system is composed only of test f\/ields in static f\/lat
space-time backgrounds, there is no need to consider the
back-reaction of the f\/ields on the geometry.

We present in section~\ref{sol} an {\em exact} solution 
of the evolution equation for the harmonic oscillator 
in f\/lat space-time whose $t=const\,$ spacelike
sections are any multiply-connected f\/lat 3-manifold.
Through a heuristic analytical reasoning we also show in
section~\ref{sol} that this exact solution $Q(t)$ 
and the corresponding energy of the harmonic oscillator $E(t)$
both exhibit a divergent {\em asymptotical\/} exponential 
behavior. Clearly the energy can be limitlessly ``extracted'' 
from the interaction term in the multiply-connected cases 
only because the system permits this extraction, since it 
is not bounded from below. But as we shall discuss in this 
work it is indeed the topology that ``excites'' this 
available ``physical mode'' of our system. 
The balance and conservation of the total energy 
of our system is also discussed in section~\ref{sol}, where we 
show that it is f\/inite and conserved for all f\/inite 
time $t>0$.

Explicit exact solution of the evolution equation for the case 
where ${\cal M}_3 = {\cal R}^2 \times {\cal S}^1 $ is discussed 
in details in section~\ref{case}. For this special case we also 
present and analyze graphs which reveal how the amplitude
$Q(t)$ and the energy $E(t)$ of the {\em overdamped\/}
harmonic oscillator vary with the time. 
The graph of $E(t)$ presents a reverberation 
pattern with relative minima and maxima, and discontinuities 
of the derivative $\dot{E}(t)\,$. The topological origin of the 
reverberations for the overdamped oscillator is also made clear 
and stressed in that section. The graphs in this section also 
conf\/irm (within the limit of accuracy of the plots, of course) 
the divergent asymptotical exponential behavior 
for both $Q(t)$ and $E(t)$, which is shown to take place 
in section~\ref{sol}.
These results make apparent that, contrarily to what one might
infer from~\cite{BernuiGomeroReboucasTeixeira98}, it is not  
necessary to have f\/lat space-time manifolds with (totally) 
compact spacelike sections ${\cal M}_3$ for the topological 
induction of the reverberations and (or) the divergent 
behavior of the energy $E(t)$ ---~the compactness 
of ${\cal M}_3$ in just one direction suf\/f\/ices.

We begin the section~\ref{finals} by emphasizing the 
importance of topological considerations in physical problems, 
and give an example that makes explicit how the overall electric
charge of universes whose $t=const$ spacelike sections are
orientable and compact f\/lat 3-manifolds is related to the 
topology. We also show that topological fragility can arise
not only in the usual cosmological modelling, but also in 
the ordinary multiply-connected f\/lat space-time manifolds we
consider in the present paper. There we stress the origin 
of the topological reverberations we found, and show that it 
constitutes an example of topological fragility, which 
takes place without violation of any local physical law. 
The extent to which this paper generalizes 
ref.~\cite{BernuiGomeroReboucasTeixeira98} is discussed in
details in section~\ref{finals}. Finally the relation of
our results with cosmology is also addressed.

%%%%%%%%
\section{Radiation Reaction Equation}  
\label{RadReacEq}
\setcounter{equation}{0} 

The f\/lat space-time manifolds we shall be concerned with are 
decomposable into ${\cal M}_4 = {\cal R} \times {\cal M}_3\,$, 
where the real line ${\cal R}$ represents a well def\/ined global 
time of Minkowski space-time.
Corresponding to the possible topologically distinct 
3-manifolds ${\cal M}_3$ there exists a simply-connected 
covering manifold ${\cal R}^3\,$, such that each manifold 
${\cal M}_3\,$ is obtained from ${\cal R}^3 $ by 
identifying points which are equivalent under the action of 
a discrete subgroup of isometries of the Euclidean
space ${\cal R}^3$ without f\/ixed points.
In other words, each manifold ${\cal M}_3 $ is obtained 
by forming the quotient space 
${\cal M}_3 = {\cal R}^3 / \Gamma$, where $\Gamma$ is a 
discrete group of isometries of ${\cal R}^3 $ acting freely 
on ${\cal R}^3\,$ and referred to as the covering group 
of ${\cal M}_3$ (for a complete classif\/ication of 
Euclidean 3-manifolds see Wolf~\cite{Wolf67}).

The physical system we shall be concerned throughout this
paper is represented by a pointlike harmonic oscillator coupled 
with a scalar f\/ield. This system has been treated by many people, 
since the work by Schwalb and Thirring~\cite{SchwalbThirring64} 
in 1964, as a simplif\/ied model to study a few features of oscillating 
(radiating) phenomena~\cite{SchwalbThirring64}~--~\cite{Bernui94}.
When the underlying manifold is ${\cal R}^3$ this system may 
serve as a model for an electric dipole coupled to electromagnetic 
radiation or an impurity atom interacting with acoustical waves~%
\cite{SchwalbThirring64}. It has also been used as simple model to 
parallel classical electrodynamics in the dipole approximation as 
discussed by Kampen~\cite{Kampen51} and Kramers~\cite{Kramers56}. It 
is worth mentioning that similar systems have also been used to mimic 
the basic properties, and thus to study the most relevant features of 
pulsating stellar systems~\cite{Schutz84}~--~\cite{Kokkotas86}. 
The essential idea that permeates these latter papers is to work 
with simple models of oscillating f\/ields interacting with oscillating 
sources in order to develop some analytical physical background to 
understand the relationship between gravitational waves and 
their sources (radiation emission and radiation reaction).
    
Consider that the harmonic oscillator of our physical 
system is located at an arbitrary point $q \in {\cal M}_3$. 
Without loss of generality, one can assume that the covering 
map $\pi:~{\cal R}^3~\rightarrow~{\cal M}_3$ maps the 
origin to $q\,$, namely $\,\pi(0,0,0) = q$. In what 
follows we shall denote by ${\cal O}_q$ the orbit  
$\pi^{-1}(q)$ of $(0,0,0)$ under the action of $\Gamma$ on 
${\cal R}^3$ (see f\/ig.~1). 

Now, the time evolution of the harmonic oscillator at 
$q \in {\cal M}_3$ can be obtained by studying the 
equivalent system on the covering manifold, which is 
formed by an inf\/inite set of indistinguishable harmonic 
oscillators each one located at a point of the orbit ${\cal O}_q\,$, 
subject to identical sets of initial conditions and interacting 
with the scalar f\/ield $\varphi\,$ in the same way as that 
of the original oscillator at $q \in {\cal M}_3$ (see f\/ig.~1).
Thus, the dynamics of our physical system in ${\cal M}_3$ can 
be derived from the following functional action for the equivalent 
system in the universal covering space ${\cal R}^3\,$:
\begin{equation} \label{act}
S = S_f + S_o + S_i \; ,
\end{equation}
where $S_f\,$ is a scalar f\/ield term, $S_o\,$ corresponds to 
the oscillators, and  $S_i\,$ is a coupling term between the 
scalar f\/ield and each harmonic oscillator. These three terms 
are, respectively,  given by
\begin{eqnarray}
S_f &=& \frac{1}{2} \int d^4x \,\, \eta^{\mu \nu}\,  
  \partial_{\mu}\,\varphi(t,\vec{x}) \,\, 
  \partial_{\nu}\,\varphi(t,\vec{x}) \;,\label{Sf} \\ 
S_o &=& \frac{1}{2} \sum_{\vec{p} \,\in\,{\cal O}_q}\, \int dt\,\, 
          [\,\dot{Q}_{\vec{p}}^2\,(t) - \omega^2 Q_{\vec{p}}^2\,(t)\,]
                                \;,\label{So} \\
S_i &=& \lambda \sum_{\vec{p} \,\in\, {\cal O}_q}\, \int d^4x \,\, 
    \rho_{\vec{p}}\,(\vec{x})\,\, \theta\,(t)\,\, 
    \varphi(t,\vec{x})\,\, Q_{\vec{p}}\,(t) \;,\label{Si}
\end{eqnarray} 
where $t \in (0, \infty)$,  
$\varphi(t,\vec{x}) = \varphi(t,\gamma\,\vec{x})$ 
is the massless scalar f\/ield, 
$Q_{\vec{p}}\,(t)$ is the amplitude of the oscillator 
located at $\vec{p} \in  {\cal O}_q$, overdot represents 
derivative with respect to the time $t$, $\,\omega$ is the 
angular frequency of each oscillator and $\lambda \neq 0$ is 
the coupling constant. Hereafter, $\rho$ is a  non-negative 
normalized density function ($\int \rho = 1$) with compact 
support (corresponding to the region of interaction) contained 
in the interior of a fundamental domain of ${\cal M}_3$ and 
centered at the origin, and f\/inally 
$\rho_{\vec{p}}\,(\vec{x}) = \rho\,(\gamma^{-1} \vec{x})$, 
where  $\gamma \in \Gamma$ and $\gamma(0,0,0) = \vec{p}\,$ 
[note that $ \rho_{\vec{0}}\, (\vec{x}) = \rho(\vec{x})\,$].
The step function $\theta\,(t)$ is used to indicate that the 
interaction starts at $t=0$. It should be noticed that the 
interaction term~(\ref{Si}) clearly is not bounded from 
below, potentially permitting a limitless extraction of 
energy by the oscillator. However, as we shall discuss later
in this article, it is the topology of the 3-space that 
``excites'' this physically available mode of the system.

A word of clarif\/ication is in order here. 
In dealing with the behavior of physical f\/ields 
in the curved space-times of general relativity one has to 
consider that these f\/ields are not only inf\/luenced by, 
but they also have gravitational ef\/fects, i.e., they 
change the geometry. 
In most ordinary situations, however, these back-reaction 
ef\/fects are small enough that can be neglected. In such cases the 
f\/ields are treated as {\em test\/} f\/ields in the corresponding 
curved space-time background. 
Had we considered our f\/ields $Q(t)$ and $\varphi(t,\vec{x})$
as {\em non-test\/} f\/ields in a {\em non-static curved\/} space-time 
we would have to take into account the back-reaction of the f\/ields 
on the geometry, and so besides $S_i$ a new interaction term would have 
to be added to the action~(\ref{act}). However, our backgrounds are 
{\em static flat\/} space-times (endowed with any one of the 17 possible 
non-trivial f\/lat topologies), and our system contains only 
{\em test\/} f\/ields with a mutual interaction, thus there is 
no place for such a term in the action of our system.  

Varying the action~(\ref{act}) with respect to $\varphi$ and
$Q$ one obtains the coupled equations of motion of the 
equivalent system, namely
\begin{eqnarray} \label{eqmot1}      
\Box \,\,\varphi(t, \vec{x}) &=& \lambda\, \sum_{\vec{p} \,\in\, 
   {\cal O}_q} \rho_{\vec{p}}\,(\vec{x})\,\,\theta\,(t)\,\,
Q_{\vec{p}}\,(t) \;, \\
\label{eqmot2}
\ddot{Q}_{\vec{p}}\,(t) + \omega^2\,\, Q_{\vec{p}}\,(t) &=& \lambda 
\int d^3x \, \rho_{\vec{p}}\,(\vec{x})\,\, \theta\,(t)\,\, 
\varphi(t,\vec{x}) \;,
\end{eqnarray}
where $\Box$ denotes the  d'Alem\-ber\-tian operator on ${\cal R}^3$.
Note that the sum in~(\ref{eqmot1}) is not an inf\/inite sum.
Actually owing to the disjoint compact supports of the functions 
$\rho_{\vec{p}}\,(\vec{x})$ the right-hand side of that equation
contains at most one summand which does not vanish.

To obtain the radiation reaction equation from the above 
equations~(\ref{eqmot1}) and~(\ref{eqmot2}) we recall that 
the general solution of eq.~(\ref{eqmot1}), considered as
an initial value problem, can be written in 
the form $\varphi(t,\vec{x})=\varphi^{}_{I}(t,\vec{x})
+\varphi^{}_{H}(t,\vec{x})$, where 
$\varphi^{}_{H}(t,\vec{x})$ satisf\/ies the corresponding 
homogeneous equation, and where $\varphi^{}_{I}(t,\vec{x})$
indicates the solution of the inhomogeneous equation. 
{}From now on we shall assume, for simplicity, that
the whole energy of the system is initially stored  
in the oscillator. Accordingly, we shall use as initial 
conditions for the equations~(\ref{eqmot1}) and~(\ref{eqmot2}) 
that $\varphi(0,\vec{x}) = \dot{\varphi}(0,\vec{x}) = 0$ and 
that $Q(0)=\alpha\,$,  $\dot{Q}(0)=\beta$, where  $\alpha$ 
and $\beta$ are real arbitrary constants.
These conditions imply that $\varphi^{}_{H}(t,\vec{x})=0$, 
and therefore $\varphi(t,\vec{x})=\varphi^{}_{I}(t,\vec{x})$. 

Now since the interaction begins at time $t=0$, the Green
function for the d'Alem\-ber\-tian operator in~(\ref{eqmot1}) 
is clearly given by%
\footnote{Clearly there are two equivalent approaches to build 
the Green function of the wave operator, which depend on 
whether it is considered on the quotient manifold 
${\cal M}_3= {\cal R}^3/\,\Gamma$, or on the covering space 
${\cal R}^3$. On ${\cal M}_3\,$, as it has been considered 
in~\cite{BernuiGomeroReboucasTeixeira98}, it has to be invariant 
under the covering transformations. On ${\cal R}^3$, however, 
which we are considering in this section, the invariance under 
$\Gamma$ must not be imposed. Note additionally that 
the covering transformations have been taken into account
in eqs.~(\ref{So})~--~(\ref{eqmot2}) and in the subsequent 
equations~(\ref{setrreq})~--~(\ref{rreq2}).}
\begin{eqnarray}
G(t,\vec{x}\,;\tau,\vec{y}\,) = \,\frac{\delta\,(t - \tau - 
|\vec{x} - \vec{y}\,|)}{4 \pi\, |\vec{x} - \vec{y}\,|} \;.
\end{eqnarray}
Hence,  the solution for equation~(\ref{eqmot1}) can be 
formally written as
\begin{equation} \label{solphi}
\varphi(t,\vec{x}) = \frac{\lambda}{4 \pi} \sum_{\vec{r}\,\in\, 
{\cal O}_q}{\int{d^3y \,\, \frac{\rho_{\vec{r}}\,(\vec{y})}{| 
\vec{x} - \vec{y} \, |}\,\, \theta\,(t - | \vec{x} - \vec{y}\, |)
\,\, Q_{\vec{r}}\,(t - | \vec{x} - \vec{y} \, |)}} \;,
\end{equation}
where we have used $\vec{r}$ as  dummy index instead of
$\vec{p}\,$.  

Finally, inserting~(\ref{solphi}) into equation~(\ref{eqmot2})
one easily obtains the radiation reaction equations for the 
oscillators, namely
\begin{equation} \label{setrreq}
\ddot{Q}_{\vec{p}}\,(t) + \omega^2\, Q_{\vec{p}}\,(t) =
\frac{\lambda^2}{4 \pi} 
\sum_{\vec{r}\,\in\,{\cal O}_q}{\int{d^3x \, d^3y \,\, 
\frac{\rho_{\vec{p}}\,\,(\vec{x})\,\, \rho_{\vec{r}}\,(\vec{y})}{| 
\vec{x} - \vec{y}\, |}\, \theta\,(t - | \vec{x} - \vec{y}\, |)
\, \, Q_{\vec{r}}\,(t - | \vec{x} - \vec{y}\, |)}} \;.
\end{equation}
As the inf\/inite set of identical harmonic oscillators 
(each one located at a point of ${\cal O}_q$) are subject 
to the same set of physical constraints, the amplitudes 
$Q_{\vec{p}}$ must evolve identically. Thus, the set
of radiation reaction equations~(\ref{setrreq})
reduces to just one dif\/ferential equation, namely
\begin{equation} \label{rreq}
\ddot{Q}\,(t) + \omega^2\, Q\,(t) =
\frac{\lambda^2}{4 \pi} \sum_{\vec{r}\,\in\,{\cal O}_q}
{\int{d^3x \, d^3y \,\, \frac{\rho\,(\vec{x})\,\, 
\rho_{\vec{r}}\,(\vec{y})}{| \vec{x} - \vec{y}\, |}\,\,\, 
\theta\,(t - | \vec{x} - \vec{y}\, |)\, \,Q\,
(t - | \vec{x} - \vec{y}\, |)}} \,,
\end{equation}
which holds for each oscillator in ${\cal O}_q \subset 
{\cal R}^3\,$, and whose solution $Q(t)$ gives the time
evolution for the amplitude of the harmonic oscillator
in ${\cal M}_3$. 

One might think at f\/irst sight that the sums on the right
hand side of eqs.~(\ref{solphi})~--~(\ref{rreq}) are 
divergent, and a regularisation scheme is needed to cope with
the divergences. However, this is not true because for any 
f\/inite time $t > 0$ the sums in those equations are in fact 
f\/inite owing to the disjoint compact supports of the functions 
$\rho_{\vec{p}}\,(\vec{x})$ together with the cut-of\/f 
ef\/fects of the step function $\theta(t)$. In other words,
we do not have formal series to be carefully handled, but
f\/inite expressions.

To deal with the pointlike coupling between the harmonic
oscillator and the scalar f\/ield [$\,\rho\,(\vec{x}) 
\rightarrow \delta^3(\vec{x})\,$], we shall consider
an inf\/inite family of oscillator--f\/ield systems
($n$-system, for short), 
with the same coupling  constant $\lambda \neq 0\,$, and 
such that each element of the $n$-system is characterized 
by a {\em bare} angular frequency $\omega_n$ and density 
function $\rho_n$, satisfying the generalized Aichelburg-Beig 
condition~\cite{AichelburgBeig76}, namely 
\begin{equation}
\label{dampcond}
\omega_n^2 - \frac{\lambda^2}{4 \pi} \int{d^3x \, d^3y \, 
\frac{\rho_n(\vec{x}) \rho_n(\vec{y}\,)}{| \vec{x} - \vec{y}\, 
|}} \equiv \Omega^2 > 0 \;,
\end{equation}
where $\Omega$ is the same constant for each member of the
$n$-system, and $\{\rho_n\}$ is a $\delta$-sequence in 
the sense of distributions~\cite{Richtmyer78}.
It should be noticed that the inequality~(\ref{dampcond}) 
is a suf\/f\/icient condition for having 
radiation damping of the energy $E(t)$ in the pointlike 
coupling limit case when the oscillator-f\/ield 
system lies in the Minkowski space-time manifold
${\cal M}_4 = {\cal R} \times {\cal R}^3\,$ 
(see, for example, ref.\ \cite{AichelburgBeig76}). 

Now, since eq.~(\ref{rreq}) holds for each member of 
the $n$-system, using relation~(\ref{dampcond}) 
one f\/inds
\begin{eqnarray} \label{rreq1}   
\lefteqn{\ddot{Q}_n(t) + 2\, \Gamma \int d^3x \,d^3y\, 
\rho_n(\vec{x}) \, \rho_n(\vec{y}) 
\;\; \frac{Q_n(t) - \theta\,(\tau) \,\, Q_n\,(\tau)} 
{| \vec{x} - \vec{y}\, |} \,} \nonumber
 \\  & &  
\; + \;\,\Omega^2 \, Q_n\,(t)\,=\,2\,\Gamma \sum_{\vec{r} 
\,\in\, \widetilde{{\cal O}}_q} \int d^3x \, d^3y \,\, 
\frac{\rho_n(\vec{x})\;\;\rho_{n,\vec{r}}\,(\vec{y})} 
{| \vec{x} - \vec{y}\, |}\;\; \theta\,(\tau)\,\, Q_n\,(\tau)\;, 
\end{eqnarray} 
where we have set $\tau = t - | \vec{x} - \vec{y}\, |$, 
$2\, \Gamma = \lambda^2/4 \pi$ and $\widetilde{{\cal O}}_q 
= {\cal O}_q - \{(0,0,0)\}$. 

In the limit $n \rightarrow \infty$  (see Appendix) we have 
that the following equation holds for all $t > 0\,$:
\begin{equation}   \label{rreq2}
\ddot{Q}\,(t) + 2 \,\Gamma \,\dot{Q}\,(t) + \Omega^2\, Q\,(t) = 
\sum_{k=1}^\infty C_k \;\, \theta(t-t_k) \;\, Q(t-t_k) \; ,
\end{equation}
where $C_k = 2\, \Gamma D_k / t_k$, $D_k$ is the number of 
points $\vec{r} \in \widetilde{{\cal O}}_q$ such that 
$|\vec{r}| = t_k$, the sequence $\{t_k\}$ is ordered 
by increasing values, and clearly $t_k \rightarrow 
\infty$ when $ k \rightarrow \infty$. 

It should be noticed that the right-hand side of the radiation
reaction equation~(\ref{rreq2}) formally contains an inf\/inite 
countable number of retarded terms of topological origin.
However, due to the $\theta$ function, at any given f\/inite time 
$t$ only a f\/inite number of terms will contribute to the sum  
of~(\ref{rreq2}).
These terms are dif\/ferent for distinct multiply-connectedness 
of the 3-manifolds ${\cal M}_3$, and vanish for the simply-connected 
manifold ${\cal R}^3\,$.
Loosely speaking they give rise to ``ref\/lected'' waves 
from the ``topological walls'', and thus account for both the 
extrema of $E(t)$ and the discontinuities of $\dot{E}(t)\,$. 
In other words they are responsible for the topological 
reverberations. 
Clearly for the simply-connected manifold ${\cal R}^3\,$, 
there is an exponential decay of $E(t)$ and no reverberation 
takes place in this case.

%%%%%%%%
\section{Solution and Asymptotical Behavior} 
\label{sol}
\setcounter{equation}{0} 

In this section we shall present an exact recursive
solution of the radiation reaction equation%
~(\ref{rreq2}), which holds for any multiply-connected 
f\/lat manifold ${\cal M}_3$.
To this end, we f\/irst consider the dif\/ferential 
operator
\begin{equation}  \label{opeq1}
D = d^2 + 2\, \Gamma d + \Omega^2 \;,
\end{equation}
which acts on real-valued piecewise smooth functions $f(t)$ 
according to
\begin{equation} \label{opeq2}
D f\, = \, \ddot{f} + 2\,\Gamma \dot{f} + \Omega^2\,f \; .
\end{equation}
It is straightforward to show that if $g(t)$ is some 
integrable function, then the solution of the equation
\begin{equation} \label{Deq}
D f\, = \,g
\end{equation}
with initial conditions $f(0)=\dot{f}(0)=0\,$ is
\begin{equation}     \label{nhomsol}
f(t) =  e^{-\Gamma t} 
\int_0^t e^{\Gamma \tau} \; h\,(t-\tau) \; 
g(\tau) \; d \tau \;, 
\end{equation}
with $\,\mu^2 = \Omega^2 - \Gamma^2\,$ and
\begin{eqnarray}     \label{hfun}
h\,(t) = \left\{ \begin{array}
{l@{\qquad \mbox{if} \qquad}l}
\mu^{-1}\,\sin\mu\,t  & \mu^2 >0 \;, \\
 t                   & \mu^2 =0 \;, \\
\nu^{-1}\,\sinh\nu\,t  & \mu^2= - \nu^2 < 0 \;.
\end{array} \right.
\end{eqnarray}
In what follows we shall discuss in detail how to f\/ind 
the solution of~(\ref{rreq2}) for the underdamped case, 
in which $\mu^2 > 0$.
The solution for the remaining two cases, however, 
can be similarly worked out, and for the sake of brevity
we will present later only the f\/inal results. 

Our purpose now is to f\/ind a real-valued function 
$\,Q: [\,0,\infty\,) \rightarrow {\cal R}\,$ of class $C^1\,$ 
such that $Q(0)=\alpha$ and $\dot{Q}(0)=\beta$ 
and satisfying~(\ref{rreq2}), which in terms of the 
operator $D$ can be rewritten as
\begin{equation}     \label{rreq3}
D\,Q(t)\, = \,\sum_{k=1}^\infty C_k \; \theta(t-t_k) 
               \; Q(t-t_k) \; ,
\end{equation}  
where the $C_k$ are positive real numbers and $\{t_k\}$ is a 
strictly increasing sequence of positive real numbers 
($0 < t_1 < t_2 < \cdots < t_k$), 
such that $t_k \rightarrow \infty$ when 
$ k \rightarrow \infty$. 

Thus, for $f(t)= Q(t) - s_0(t)$ and taking into 
account~(\ref{Deq}) and (\ref{nhomsol}) the solution 
of (\ref{rreq3}) can be written as
\begin{equation}  \label{impsol1}
Q(t) = s_0(t) + \frac{1}{\mu} \; \sum_{k=1}^\infty C_k \; 
\theta(t-t_k) \;e^{-\Gamma (t-t_k)} \int_0^{t-t_k} 
e^{\Gamma \tau} \;\sin \mu (t-t_k-\tau) \; Q(\tau) \; 
d \tau \; ,
\end{equation}
where the function $s_0$ must satisfy the homogeneous
equation $D\,s_0(t) = 0$ with the conditions $s_0(0)=\alpha\,$, 
$\dot{s}_0(0)=\beta$. Clearly one can rewrite the 
solution~(\ref{impsol1}) in the form
\begin{equation}      \label{impsol2}
Q(t) = s_0(t) + \sum_{k=1}^\infty C_k \; \theta(t-t_k) \;
Q_1\,(t-t_k) \; ,
\end{equation}
where, from~(\ref{Deq}) and (\ref{nhomsol}), the function
\begin{equation}  \label{Q1eq}
Q_1\,(t) = \frac{1}{\mu}\; e^{-\Gamma t} \int_0^{t} 
e^{\Gamma \tau} \;\sin \mu (t-\tau) \; Q(\tau) \; 
d \tau 
\end{equation}
fulf\/ils the dif\/ferential equation $D\,Q_1\,(t) = Q(t)\,$ with 
the initial conditions $\,Q_1\,(0)=\dot{Q}_1(0)=~0\,$.

The solution~(\ref{impsol2}) is given in an implicit form.
An explicit form can be obtained, though.
Indeed, as we are concerned only with real-valued functions 
def\/ined for $t \in [0,\infty)\,$, if we recursively 
def\/ine functions $s_n\,$ by
\begin{eqnarray} 
D\,s_n(t) &=& s_{n-1}(t) \quad \mbox{for} \quad n \geq 1 
                   \label{impS} \;, \\
D\,s_0(t) &=& 0 \label{impSn1} \;,
\end{eqnarray}
with the conditions $s_n(0)=\dot{s}_n(0)=0\,$ for 
$n \geq 1\,$, and functions $Q_n(t)\,$ by 
\begin{eqnarray} 
D\,Q_n\,(t) &=& Q_{n-1}(t) \quad \mbox{for} \quad n \geq 1 \;, 
                    \label{impQ} \\
Q_0\,(t)    &=& Q(t) \; , \label{impQn1}
\end{eqnarray}
with $Q_n\,(0)=\dot{Q}_n(0)=0\,$ for $\,n \geq 1\,$, then  
we can show by induction that
\begin{equation} \label{induction} 
Q_n\,(t) = s_n(t) + \sum_{k=1}^\infty C_k \; \theta(t-t_k) \;
Q_{n+1}\,(t-t_k) \, .
\end{equation}
Notice that the f\/irst of these equations (for $n=0$)
is nothing but equation~(\ref{impsol2}). 

An explicit form for the solution of~(\ref{rreq3}) can
now be obtained as follows. Equations~(\ref{impS})
and~(\ref{impSn1}) can be integrated to give
\begin{eqnarray} 
s_0(t) & = & e^{-\Gamma t} \; (A \sin\mu t + B \cos \mu t) 
\; , \label{expS} \\
s_n(t) & = & \frac{1}{\mu} \; e^{-\Gamma t} \int_0^t 
e^{\Gamma \tau} \sin \mu (t-\tau) \; s_{n-1}(\tau) \; d \tau 
\quad \mbox{for} \quad n \geq 1 \; , \label{expSn1} 
\end{eqnarray}
where $A = (\beta + \alpha\,\Gamma)\, / \, \mu \,$, $B=\alpha\,$,
and where obviously we have used~(\ref{opeq2})~--~(\ref{hfun}).
On the other hand, if we def\/ine
\begin{eqnarray} 
q_0(t)  &= & s_0(t)  \; , \nonumber  \\
q_1(t)  &= & \sum_{k_1}  \; C_{k_1} \; \theta(t - t_{k_1}) \; 
            s_1(t - t_{k_1}) \;, \nonumber \\  
q_2(t)  &= & \sum_{k_1,k_2}  \; C_{k_1}\,C_{k_2} \; 
          \theta(t-t_{k_1}-t_{k_2}) \; 
            s_2(t - t_{k_1}- t_{k_2}) \;, \nonumber \\
                  & \vdots &  \label{qieq}   \\
q_i(t)  &= & \sum_{k_1,k_2,\ldots,k_i}  \; \left(\prod_{j=1}^i 
C_{k_j}\right) \; \theta(t - \sum_{j=1}^i t_{k_j}) \; 
s_i(t - \sum_{j=1}^i t_{k_j}) \;,  \nonumber
\end{eqnarray}
then the explicit form of the solution of equation~(\ref{rreq3}) 
is simply given by
\begin{eqnarray}        \label{expsol1}
\hspace{-4cm}\lefteqn{Q(t) = \sum_{i=0}^\infty\;\, q_i(t) \; . }
\end{eqnarray}

It should be noticed that since the above solution 
is ultimately given in terms of $s_0\,$,
equations~(\ref{expS})~--~(\ref{expsol1}) make apparent 
that $Q(t)$ is completely determined by the pair of 
initial conditions  
[$\,Q(0)=\alpha\:,\:\dot{Q}(0)=\beta$\,] as one would
have expected from the outset.

Another important point to be stressed regarding 
the above solution is that the right-hand side 
of~(\ref{expsol1}) is not an inf\/inite sum as 
it appears at f\/irst sight. Actually, for a
given f\/inite time $t\,$ the above sum will clearly
be truncated by the presence of the step function 
[$\,\theta(t) = 0$ for $t \leq 0\,$, $\theta(t) = 1$ 
for $t > 0\,$], i.e. only those $q_i(t)$ 
with $i < t\,/\,t_1\,$ will contribute to the 
sum~(\ref{expsol1}). 
Moreover, each $q_i(t)$ as given by~(\ref{qieq}) is
itself  a f\/inite sum, again due to the cut-of\/f ef\/fects 
of the step function $\theta(t)\,$.

Before proceeding to the discussion of the asymptotical 
behavior of $Q(t)$ we mention that the solution
of~(\ref{rreq2}) for the critically damped ($\mu^2= 0\,$)
and overdamped ($\mu^2 = -\nu^2 < 0$) cases is again 
given by~(\ref{expsol1}) with (\ref{qieq}), but now with 
the functions $s_0$ and $s_n$ given by 
\begin{eqnarray} 
s_0(t) & = & e^{-\Gamma t} \; (A\,+ B\,t)  
      \qquad \; \mbox{for} \qquad  \mu^2 = 0  \label{scd} \;,  \\
s_0(t) & = & e^{-\Gamma t} \;(A \sinh\nu t + B \cosh \nu t)
     \quad \mbox{for} \quad  \mu^2 = - \nu^2 < 0 \label{sod} \;,\\
s_n(t) & = & e^{-\Gamma t} \int_0^t 
e^{\Gamma \tau}\; h\,(t-\tau) \; s_{n-1}(\tau) \; d \tau 
\quad \mbox{for} \quad n \geq 1 \label{sncdod} \;,
\end{eqnarray} 
where $A$ and $B$ are real constants, which are determined by
the initial conditions, and the functions $h(t)$ are given by 
equations~(\ref{hfun}).

Now, for any non-trivial f\/lat topology the asymptotic 
behavior of the solution $Q(t)$ of equation~(\ref{rreq3}) 
can be f\/igured out by a procedure similar to that used in%
~\cite{BernuiGomeroReboucasTeixeira98}.  
Indeed, if one {\em ad hoc\/} assumes an exponential 
asymptotical behavior for $Q$ of the form
$Q(t)= \gamma\, \,\, \mbox{exp}\,(\sigma \, t)$ with
$\sigma$ and $\gamma$ real constants, then 
in the limit $t \rightarrow \infty$ equation~(\ref{rreq3}) 
reduces to
\begin{equation} \label{asingen}
\sigma^2 + 2\, \Gamma\, \sigma + \Omega^2 = 
\sum_{k=1}^\infty C_k\; e^{-\,\sigma\, t_k} \; .
\end{equation}
This equation has only one real solution  for
$\sigma$. Indeed, the right-hand side of%
~(\ref{asingen}) is a positive monotone 
decreasing function $r\,(\sigma)$ with 
$\lim_{\sigma \to  0} \: r\,(\sigma) = \infty\,$ and
$\lim_{\sigma \to  \infty } \: r\,(\sigma) = 0 \,$. 
Thus $r\,(\sigma)$ lies entirely in the f\/irst quadrant 
of the plane and crosses it from the top-left to the 
bottom-right (see f\/ig.~2).
Now, since $\Gamma > 0$ then for a given pair 
($\Gamma, \Omega$) the left-hand side of~(\ref{asingen}) 
is a parabola curved upwards with vertex at 
$\sigma = - \Gamma$. Therefore, it always intersects
the curve for $r\,(\sigma)$ in just one point, 
which is in the f\/irst quadrant (see f\/ig.~2). 
This amounts to saying that there is only one real 
number $\sigma = b > 0$ solution 
to equation~(\ref{asingen}). 
Hence the amplitude $Q(t)$, and consequently the 
energy of the oscillator 
$E(t)=\frac{1}{2}\,[\,\dot{Q}^{2}(t) +\Omega^2\,Q^{2}(t)]$, 
both exhibit an asymptotical exponentially divergent behavior
(conf\/irmed, within the limits of accuracy of the plots,
by f\/igures~3 and 4 for the special case of
an overdamped oscillator discussed below).
One can sum up by stating that the compactif\/ication 
in just one direction is suf\/f\/icient to topologically induce 
a asymptotic divergent behavior of $E(t)$.

A question which naturally arises here is whether the total
energy of our system is conserved. As our system contains only
mutually interacting test f\/ields $\,Q(t)$ and $\varphi(t,\vec{x})\,$
in static f\/lat background, the energy of the overall system must
be conserved. Indeed, for $t>0$ we have $\theta(t)=1$ and the 
action~(\ref{act})~--~(\ref{Si}) does not depend explicitly on the 
time. Thus, for any well-behaved density function $\rho(\vec{x})$ 
the total energy of our system ${\cal E}(t)$ clearly is conserved, 
which squares with the fact that our system is a test system 
in a static background. For the sake of completeness, in 
what follows we shall show in more details how the total energy 
of our system ${\cal E}(t)$ is f\/inite and conserved for all time 
$t \in (0, \infty)$ even in the pointlike limit case 
$\,\rho\,(\vec{x}) \rightarrow \delta^3(\vec{x})\,$. We shall also 
qualitatively discuss how the balance 
of ${\cal E}(t)$ can be handled.
 
{}From the functional action~(\ref{act})~--~(\ref{Si}) the energy
${\cal E}(t)$ of our system can clearly be written in the form
\begin{equation} \label{tenergy}
{\cal E}(t) = {\cal E}_f(t) + {\cal E}_{o}(t) + {\cal E}_i(t) \;,
\end{equation}
where 
\begin{eqnarray} 
{\cal E}_f(t) &=& \frac{1}{2}\,\int_{{\cal D}} d^3x\;
              \{\,{\dot{\varphi}}^2\,(t,\vec{x})+[\,\nabla 
               \varphi\,(t,\vec{x})\,]^2\,\}\;, \label{Ef} \\
{\cal E}_{o}(t) &=&\frac{1}{2}\;[\,\dot{Q}^2\,(t)
               +\omega^2\,Q^2\,(t)\,]   \;,\label{Eo} \\               
{\cal E}_i(t) &=& - \lambda\;\, Q(t) \int_{{\cal D}} d^3x \;
              \rho(\vec{x}) \, \varphi(t,\vec{x}) \;, \label{Ei}
\end{eqnarray}
for $t \in (0,\infty)$. Here ${\cal D}$ denotes a fundamental domain
of ${\cal M}_3$. As we have mentioned above for any
time $ t>0$ the energy ${\cal E}(t)$ is conserved since the 
corresponding Hamiltonian does not depend explicitly on the time. 

{}From equation~(\ref{solphi}) the interaction component 
${\cal E}_i(t)$, for which the f\/ield $\varphi(t,\vec{x})$ is 
evaluated at $\vec{x}=0$, i.e. in the location of the pointlike 
oscillator, clearly presents a divergent behavior of the form 
$|\vec{x}|^{-1}$ in the neighborhood of $\vec{x}=0$ (see below
for details). Now, to deal with the balance of the total 
energy ${\cal E}(t)$, we introduce the divergent quantity
\begin{equation} 
\label{lambd}
\Lambda = \frac{\lambda^2}{4 \pi} \;\lim_{r \rightarrow 0+}
\, \frac{1}{r}
\end{equation}
in terms of which a f\/inite energy $E_i(t)$ can be 
def\/ined for all $0<t<\infty$, namely
\begin{equation}  \label{Eif}
{\cal E}_i(t) = E_i(t) - \Lambda\;Q^2(t)\;.
\end{equation}
Thus, for each time $t$ in that interval, clearly one can extract 
from ${\cal E}_i(t)$ an inf\/inite amount of negative energy to 
obtain a regularized f\/inite negative energy $E_i(t)$.

{}The energy ${\cal E}_f(t)$ as given by~(\ref{Ef}) also diverges.
In fact, from~(\ref{solphi}) one f\/inds that in the neighborhood
of $\vec{x}=0$ the scalar f\/ield $\varphi(t,\vec{x})$ 
takes the form
\begin{equation} \label{phiori}
\varphi(t,\vec{x})=\frac{\lambda}{4 \pi} \;\frac{Q(t)}{|\vec{x}|}
                    + (n.d.) \;,
\end{equation}
where $(n.d.)$ denotes the non-diverging terms for any 
$t \in (0,\infty)$. So, in the neighborhood of $\vec{x}=0$ 
we have
$|\nabla \varphi| = |\,(\lambda / 4 \pi \,  )\; Q(t)\,/\,|\vec{x}|^2\,|$, 
and the integral corresponding to this term in eq.~(\ref{Ef}), namely
$\frac{1}{2}\,\int_{{\cal D}} 4\,\pi\,r^2\, dr\;|\nabla \varphi|^2\,$,
has a divergent contribution of the form $ \Lambda\,Q^2(t)/\,2$
when $ r = |\vec{x}| \rightarrow 0$. Now since the contribution
of the term $\dot{\varphi}^2$ to the integral on the right-hand
side of~(\ref{Ef}) is f\/inite for all $t \in (0,\infty)$, one can 
similarly introduce a regularized f\/inite energy $E_f(t)$
by
\begin{equation} \label{Eff}
{\cal E}_f(t) = E_f(t) + \frac{1}{2}\;\Lambda\;Q^2(t)\;,
\end{equation}
which is clearly f\/inite for all f\/inite time $t >0$.

The energy component corresponding to oscillator ${\cal E}_o(t)$ 
as given by~(\ref{Eo}) can be dealt with by using the 
frequency $\Omega$ given by eq.~(\ref{dampcond}) in the limit case 
$\,\rho\,(\vec{x}) \rightarrow \delta^3(\vec{x})\,$, namely
using $\omega^2 = \Omega^2 + \Lambda $. Indeed, with this
renormalized frequency one can def\/ine
\begin{equation}   \label{Eof}
{\cal E}_o(t) = E_o(t) + \frac{1}{2}\;\Lambda\;Q^2(t)\;,
\end{equation} 
where the regularized energy
$E_o(t)=\frac{1}{2}\,[\,\dot{Q}^{2}(t) +\Omega^2\,Q^{2}(t)]$
is f\/inite for all f\/inite time $t>0$. The energy of the
oscillator $E_o(t)$ has also been denoted simply by $E(t)$ 
in many places of this paper.

Finally, we note that according to eqs.~(\ref{Eif}), (\ref{Eff}) 
and~(\ref{Eof}), the divergent terms of the form $\Lambda\, Q^2(t)$ 
cancel out in the sum ${\cal E}_i (t)+{\cal E}_f(t)+{\cal E}_o(t)$. 
Thus, from~(\ref{tenergy}) one obtains that the total energy of the 
system ${\cal E}(t)$ given by~(\ref{tenergy}) reduces to
\begin{equation}
{\cal E}(t) =  E_f(t) + E_o(t) + E_i(t)\;,
\end{equation}
which is constant and f\/inite for all time $t>0$.

The fact that the total energy of the system ${\cal E}$ 
is f\/inite and conserved for all time $t>0$ is undoubtedly
the most important physical requisite our system obeys. 
However, it is the fact that the interaction energy is not 
bounded from below which makes possible and understandable the 
asymptotical divergent behavior of the energy of the oscillator 
and of the scalar f\/ield, since they both can ``extract'' energy 
from the interaction term. Remarkably this latent degree of freedom 
of our system is excited only when it lies in the multiply-connected 
manifolds, since when it is in ${\cal R}^3$ there is a damping 
of energy of the oscillator. In other words, the topology 
(multiply-connectedness) excites this available ``physical mode'' 
of our system. This remarkable feature of this system has not been 
perceived since the sixties, when it was f\/irst examined by 
Schwalb and Thirring (see, for example, refs.%
~\cite{SchwalbThirring64}~--~\cite{HoenselaersSchmidty89}).

It is certainly desirable that quantum physical systems have 
the interaction energy bounded from below. However, at a classical 
level this is not an imperative. When one considers, 
for example, a system of two isolated pointlike charges with opposite 
signs in a f\/lat space (vacuum medium), the smaller is the distance 
$d$ between the charges the greater is the absolute value of potential 
energy of the system. Even if one assumes that the charges have 
already been renormalized, in the limit $d \to 0$ the potential energy 
formally diverges. This sort of divergences are formal, since the 
limit $d \to 0$ cannot be attained in practice. Furthermore, one
ought to bear in mind that the classical theory used to infer 
this divergence does not hold for every small $d$ (compared with, 
e.g., the Bohr radius). Actually, the distance $d$ is assumed
to be larger than the Bohr radius since otherwise the self-f\/ields
of the two interacting charges would be rather distorted, and one 
should expect much more involved physical ef\/fects.
Similarly in our system the energy of the oscillator $E_{o}(t)$ is 
f\/inite for all f\/inite $t > 0$, but formally diverges in the limit 
$t \to \infty$. This limit case may perhaps be excluded on 
physical grounds. However, the most remarkable point here is 
that a well-behaved physical system when it lies in the ordinary 
simply-connected Euclidean manifold ${\cal R}^3$, will exhibit 
an unexpected behaviour (reverberation pattern followed by a 
growth of $E(t)$) when it is  in any possible flat manifold with 
non-trivial topology, without violating any local physical law. 

%%%%%%%%
\section{Case Study}
\label{case}
\setcounter{equation}{0} 

In this section we shall focus our attention on the time 
evolution of the harmonic oscillator in a f\/lat space-time 
whose $t=const$ spacelike section is the  
simplest multiply-connected Euclidean 3-manifold, 
namely ${\cal M}_3 = {\cal R}^2 \times {\cal S}^1\,$.
In this case, denoting by  $a$ the distance between 
two equivalent  points in the covering manifold 
${\cal R}^3$, one can easily f\/ind that
$C_k = 4\, \Gamma/t_k\,$, where $t_k = k\,a\,$ ($k$ stands for
a positive integer).
Hence, from (\ref{qieq}) and for $i>0\,$ one 
obtains the solution 
\begin{equation} \label{q_cilinder}
q_i(t) = \left(\frac{4 \Gamma}{a}\right)^i \sum_{k_1, 
k_2, \ldots, k_i} \left(\,\prod_{j=1}^i\frac{1}{k_j}\right)\; 
\theta(t - a \sum_{j=1}^i k_j) \,\; s_i(t - a \sum_{j=1}^i k_j) \;,
\end{equation}
which together with~(\ref{expsol1}) and the f\/irst eq.~(\ref{qieq})  
give the behavior of the amplitude $Q(t)$ and of the energy
$E(t)=\frac{1}{2}\,[\,\dot{Q}^{2}(t) +\Omega^2\,Q^{2}(t)]$
for the following dif\/ferent types of oscillator: (i) the
underdamped ($\,\mu^2  > 0$) for which $s_0$ and $s_n$ are given 
by~(\ref{expS}) and~(\ref{expSn1}); (ii) the critically damped 
($\mu^2= 0\,$) where~(\ref{scd}) and (\ref{sncdod}) furnish $s_0$ 
and $s_n\,$; (iii) the overdamped ($\mu^2 < 0$) for which $s_0$ 
and $s_n$ are given by~(\ref{sod}) and (\ref{sncdod}).

As far as the asymptotical behavior of $Q(t)$ for this special 
case (${\cal M}_3 = {\cal R}^2\, \times {\cal S}^1\,$)
is concerned, the right-hand side series of~(\ref{asingen}) 
can easily be evaluated and equation~(\ref{asingen}) becomes
\begin{equation} \label{asincil}
\sigma^2 + 2 \, \Gamma \,\sigma + \Omega^2 = 
-\,\frac{4\, \Gamma}{a}\, \ln\,(1-e^{-\,\sigma\, a}) \; .
\end{equation}
Clearly, for a specif\/ic oscillator, i.e. for a given pair
($\Gamma, \Omega$), this equation can be numerically solved for 
$\sigma$ by using, for example, a computer algebra system such 
as Maple~\cite{Maple,Heck}. For an overdamped oscillator with
$\Gamma=6\,$ and $\Omega =5$, which we shall use in our numeric
calculations, taking $a=1$ and using Maple one easily obtains 
the approximate solution $\sigma = b = 0.35\,$.

In the remainder of this section we shall discuss the 
f\/igures~3 and~4 and make some further remarks. 
We begin by emphasizing that we have chosen to discuss the 
dynamical behavior of both the amplitude and the energy of
the oscillator in a space-time whose spacelike section is 
${\cal M}_3 = {\cal R}^2 \times {\cal S}^1$ because this is 
the simplest multiply-connected f\/lat manifold (since it is 
compact in just one direction) having therefore the lowest 
degree of connectedness for a f\/lat 3-manifold%
~\cite{BernuiGomeroReboucasTeixeira98}. 
This amounts to saying that the dynamical ef\/fects exhibited 
by the physical quantities in this manifold will also be
present in any other multiply-connected f\/lat 3-manifold.
As a matter of fact, the greater is the degree of connectedness
of ${\cal M}_3$ the more reinforced will be these dynamical
ef\/fets of topological nature. The limiting simply-connected
case ${\cal R}^3$ will manifest no sign of such ef\/fects.
A second point of general order which is worth noting is 
that we have decided to plot the graphs for an overdamped 
oscillator because we knew from the outset that the amplitude 
$Q(t)$ and the energy $E(t)$ for this oscillator, when in 
${\cal R}^3\,$, exhibit an exponential decay with no relative 
extrema.
This choice amounts to freezing the degrees of oscillations,
and therefore rules out any sort of resonance, since clearly 
there is no frequency to be coupled with to resonate. Besides, 
our choice also makes easier and neater the comparison of 
the dynamical behavior of the oscillator in the simply and 
multiply connected cases. 
Taking into account these considerations we have plotted the 
f\/igures~3 and~4 for an overdamped harmonic oscillator with 
$\Gamma =6\,$, $\Omega = 5$ and with initial conditions
$Q(0)=1$ and $\dot{Q}(0)=0$, in a f\/lat space-time whose 
$t=const$ spacelike section ${\cal M}_3$ is endowed with the 
topology ${\cal R}^2 \times {\cal S}^1\,$, and where we have 
taken the distance $a=1$ in eqs.~(\ref{q_cilinder}) and
(\ref{asincil}). The analysis of these f\/igures is given 
in what follows.

Figure~3 shows the graph for the amplitude $Q(t)$ which exhibits
relative maxima and minima followed by a growth of $Q(t)$. 
This time evolution of the amplitude sharply contrasts 
with the behavior of the amplitude $Q(t)$ for the same 
oscillator in ${\cal R}^3$, where neither relative extrema 
nor growth of $Q(t)$ take place.
This f\/igure also contains the graph of the function 
$\exp(-bt)\,Q(t)\,$, which within the accuracy of 
numeric calculations performed for the plots 
conf\/irms an exponential form for the divergent behavior 
suggested by the graph of $Q(t)$ and shown to indeed take
place in the previous section.
In this graph we have used $b=0.35$, which is the root 
of eq.~(\ref{asincil}) for $\Gamma =6\,$,  $\Omega = 5$
and $a=1$.

Figure~4 shows the behavior of the energy $E(t)$ which 
exhibits a few relative minima and maxima followed by an 
asymptotic divergent behavior. This divergent behavior is 
clearly in agreement with the behavior of $Q(t)$ when 
$t \rightarrow~\infty\,$ shown in f\/igure~3.
The time evolution of the energy for the oscillator in 
this multiply-connected manifold is in striking contrast 
with the behavior of the energy $E(t)$ in the simply-connected 
case ${\cal R}^3$, where there is an exponential damping of 
$E(t)$ with no relative extrema. Figure~4 also contains the 
graph of the derivative $\dot{E}(t)$, exhibiting discontinuities 
at each $t=t_k=k\,a$ ($k$ stands for a positive integer),
which again do not occur in the simply-connected case 
${\cal R}^3$. The graph of $\dot{E}(t)$ also reveals that 
$E(t)$ has a f\/inite number of extrema. 

It should be also noticed that the derivative 
$\dot{E} = \dot{Q}\,(\, \ddot{Q} + \Omega^2\, Q\,)$ 
of the energy function exhibits discontinuities not only in the 
specif\/ic case handled in this section, but also in the general 
underlying topological setting of this work. Indeed, from 
equation~(\ref{rreq2}) one easily obtains
\begin{equation}
\dot{E}(t) = 2 \, \Gamma \, \dot{Q}(t) \left[ \, 
 \sum_{k=1}^\infty \frac{D_k}{t_k} \,\,
 \Theta(t-t_k) \,\, Q\,(t - t_k) \, - \,\dot{Q}\,(t) \right] \,.
\end{equation} 
{}From this equation one has that the discontinuities occur
at $t=t_k$, i.e. they come about each time a new term
$\,Q(t-t_k)\,$ is taken into acount in the right-hand side
of eq.~(\ref{rreq2}).

The relative extrema and asymptotic divergent behavior of
$E(t)$ as well as the discontinuities of $\dot{E}(t)\,$, which
occur only in the multiply-connected f\/lat manifolds, are 
due to the retarded terms of topological origin in the 
evolution equation~(\ref{rreq2}). These topological terms 
are dif\/ferent for distinct connectedness of the 
$t=const$ section ${\cal M}_3$ of the f\/lat space-time 
manifold, and  vanish for the simply-connected case 
${\cal R}^3$. They are ultimately responsible for those 
features (extrema, discontinuities and asymptotic behavior) 
of $E(t)$. In brief, as it has been made apparent for the
overdamped oscillator, both the suppression of the radiation 
damping and the reverberation pattern of $E(t)$ in those 
multiply-connected
f\/lat space-times are of purely topological origin, and 
on the other hand, evince that our system is topologically 
fragile~\cite{ReboucasTavakolTeixeira98}.

%%%%%%%%

\section{Concluding Remarks}  
\label{finals}
\setcounter{equation}{0} 

Since the physical laws are usually expressed in terms of
local dif\/ferential equations the topological considerations 
may not be prominent at f\/irst sight. Nevertheless they are often
necessary in many problems of physics. In general relativity and
cosmology, which handles the dynamic and the global structure 
of space-time manifolds, the study of topological features
is even more signif\/icant and acquires a dynamical meaning 
in a sense.

Topological considerations are necessary in many other
situations in physics. When one considers, for example, the 
electric f\/ields produced by bounded sources, one quite often 
chooses as boundary condition that the f\/ield vanishes at spatial 
inf\/inity. This choice is possible and even convenient if the 
$t=const$ spacelike section of the Minkowski space-time 
${\cal M}_4$ is the ordinary simply-connected Euclidean 
manifold ${\cal R}^3$. However, as far as multiply-connected 
spacelike three-spaces ${\cal M}_3$ are concerned the choice 
of satisfactory boundary conditions for this problem is not as 
simple as that.
So, for example, in the exam of the electric f\/ield produced
by an isolated point charge in any one of the six possible 
orientable compact (multiply-connected) topologically distinct 
Euclidean 3-manifolds ${\cal M}_3$, clearly this condition 
at inf\/inity cannot be imposed. 
Actually, it is easy to show by assuming the Gauss's law that 
one cannot have a net electric charge in these compact 
f\/lat manifolds, making clear that the overall electric 
charge in these manifolds is related to the 
topology~\cite{EllisSchreiber86}.
This example evinces that changes in the assumption of the 
space-time topology may induce important physical consequences 
even in the case of static f\/lat space-time manifolds. 
This type of sensitivity, which is also present in cosmological 
modelling~\cite{ReboucasTavakolTeixeira98}, has been referred 
to as topological fragility and can clearly occur without 
violation of any local physical law%
~\cite{BernuiGomeroReboucasTeixeira98,ReboucasTavakolTeixeira98}.
As a matter of fact, in the above example the topological fragility
arises when we impose the local validity of an ordinary law
of classical electromagnetism to a system which lies in a f\/lat
multiply-connected manifold.
  
In the same context, in this work we have studied the role 
played by multiply-connectedness in the time evolution 
of the energy $E(t)$ of a radiating system that lies in static 
f\/lat space-time  manifolds ${\cal M}_4$ whose $t=const\,$ 
spacelike sections ${\cal M\/}_3$ are compact in at least one 
spatial direction. So, it may even have the lowest degree of 
multiply-connectedness. This topological setting is general
enough to encompass the entire set of possible classes of 
multiply-connected f\/lat 3-manifolds discussed, for 
example, by Wolf~\cite{Wolf67}. 
We have shown that the behavior of the radiating energy $E(t)$
of the oscillator changes remarkably from exponential damping,
when the underlying 3-manifold is ${\cal R}^3$, to
a reverberation behavior for $E(t)$ when the 
spacelike $t=const$ sections of ${\cal M}_4$ are any possible 
multiply-connected f\/lat 3-manifold. Thus, our study
shows that the topological fragilities can arise not only
in the usual cosmological modelling~%
\cite{ReboucasTavakolTeixeira98}, but also in ordinary
static f\/lat space-time manifolds as long as they are 
multiply connected. As we have emphasized in section%
~\ref{RadReacEq} and~\ref{sol} 
this striking behavior of $E(t)$ occurs with no violation 
of any physical law, and has a purely topological origin.

Although the physical system we have investigated 
here is the same system discussed in%
~\cite{BernuiGomeroReboucasTeixeira98}, the present 
article generalizes the results of that paper in several
respects. Firstly, the underlying topological 
setting is much more general since we did not assume 
that our spacelike 3-manifolds ${\cal M}_3$ are compact 
and orientable. Actually we only require that they are
multiply-connected, which means that ${\cal M}_3$ can
be endowed with any one of the 17 multiply-connected 
f\/lat topologies~\cite{Wolf67}, including, of
course, the orientable compact 3-manifolds 
dealt with in~\cite{BernuiGomeroReboucasTeixeira98}.
Secondly, contrarily to the {\em numerical\/} integration 
performed in~\cite{BernuiGomeroReboucasTeixeira98}, 
here we have obtained a closed {\em exact\/} solution of the
evolution equation for the harmonic oscillator
in the above-mentioned general topological setting.
Thirdly, in the case study we discussed in details in 
section~\ref{case} we have found the (exact) time 
evolution of our system in a specif\/ic {\em non-compact\/} 
(although multiply-connected) manifold, while in%
~\cite{BernuiGomeroReboucasTeixeira98} the time behavior
of our system was (numerically) treated only for six 
{\em compact orientable\/} background manifolds. 
Finally in the present article we have also discussed
the conservation and the  balance of the total energy 
of our physical system, making apparent that in both papers
the total energy is f\/inite and conserved for any time 
$t>0$.

Before closing it is worth mentioning that to the
extent that we have studied the dynamical behavior
of a test system in {\em static\/} f\/lat FRW space-time 
backgrounds our results have a moderate, though    
suggestive and well-founded, relation to cosmology. 
Besides, it should be noticed that the net role played 
by topology is better singled out in this static 
case, where the dynamical degrees of freedom have been 
frozen. On the other hand, it appears that, apart from the
inf\/lationary expansion and perhaps a few other cases, the 
restriction we have made will not be decisive for 
the pattern of the behavior radiating energy $E(t)$ in a number 
of f\/lat expanding locally homogeneous and isotropic FRW 
cosmological models. Finally, it is worth noting 
that since geometry constrains, but does not dictate
the topology of space-time manifolds, in cosmological
modelling one often is confronted with the basic question 
of what topologies are physically acceptable for a given 
space-time manifold. An approach to this problem
is to study the possible physical (observational)
consequences of adopting particular topologies for
the space-time (for a fair list of references on this
approach see the review article by  Lachi\`eze-Rey 
and Luminet~\cite{LachiezeReyLuminet95}). 
This paper constitutes an example of this relevant 
approach.

%%%%%%%%
\section*{Acknowledgements}

We would like to express our thanks to Professor John A. Wheeler 
for stimulating correspondence on this subject matter.
We are also grateful to the Brazilian scientif\/ic 
agencies CAPES, FAPERJ and CNPq for the grants under
which this work was carried out.

%%%%%%%%
\appendix
\vspace{1cm}
\section*{Pointlike Coupling Limit Case}

\setcounter{equation}{0}
\renewcommand{\theequation}{A.\arabic{equation}}

Our aim in this appendix is to show how one can derive
from equation~(\ref{rreq1}), which holds for an arbitrary 
density function $\rho_n\,(\vec{x})$ of a $\delta$-sequence, 
the radiation reaction equation~(\ref{rreq2}) for the limiting 
case of pointlike coupling between the scalar f\/ield and 
the oscillator 
[$\,\rho\,(\vec{x})~\to~\delta^3(\vec{x})\,$].
 
Let us f\/irst rewrite the equation~(\ref{rreq1}) in the 
following form
\begin{equation}    \label{A1}
\ddot{Q}_n(t) + 2\,\Gamma\, I_n(t) + \Omega^2 \, Q_n(t) = 
\,2\, \Gamma \sum_{\vec{r}\, \in \,\widetilde{\cal O}_q}
 J_{n,\vec{r}}\,(t) \; ,
\end{equation}
where we have set
\begin{equation}   \label{A2}
I_n(t)  =  \int d^3x \,\, d^3y \,\, \rho_n(\vec{x})\,\, 
\rho_n(\vec{y}) \;\;
\frac{Q_n(t) - \theta(\tau) \,\, Q_n(\tau)} {| \vec{x} - 
\vec{y}\, |} \;\,,
\end{equation}
%%%%%%
\begin{equation}     \label{A3}
J_{n,\vec{r}}\,(t)  =  \int d^3x \,\, d^3y \,\, \frac{
\rho_n(\vec{x}) \,\,\rho_{n,\vec{r}}(\vec{y})}{
| \vec{x} - \vec{y}\, |}\;\, \theta(\tau)\; Q_n(\tau) \;,
\vspace{3mm}
\end{equation}
with $\tau = t - | \vec{x} - \vec{y}\, |\,$, $2\,\Gamma = 
\lambda^2/4\pi\,$ and $\,\widetilde{\cal O}_q = 
{\cal O}_q - \{(0,0,0)\}$. 

As we mentioned earlier the pointlike limit case can be 
achieved by taking the limit $ n \rightarrow \infty$ of 
the above $n$-system equation. To this end we f\/irstly 
remind that as we are dealing with density functions
$\rho_n\,(\vec{x})$ whose supports are connected sets,
the following reworded version of the mean value theorem 
can be stated: let $f(\vec{x},\vec{y})$ be an arbitrary 
continuous function def\/ined in $\,{\cal I} = \mbox{supp}
\,(\rho_n)\,\times\,\mbox{supp}\,(\rho_{n,\vec{r}})\,$, 
then there exists a point 
$(\vec{x}_0\,, \vec{y}_0) \in {\cal I}$ such that
\begin{equation}
f(\vec{x}_0, \vec{y}_0) = \int d^3x \, d^3y \,\,\rho_n\,(\vec{x}) 
\; \rho_{n,\vec{r}}\,(\vec{y})\; f(\vec{x}, \vec{y}) 
\end{equation}
holds.

Let now $R_n$ be the radius of $\mbox{supp}\,(\rho_n)\,$ and
consider a point $\vec{r}\,\in\,\widetilde{{\cal O}}_q$, 
then by the mean value theorem there exists a sequence 
of pair of points
$(\vec{x}_{0,n}\,,\, \vec{y}_{0,n}) \in {\cal I}$ such that 
for each  $t>0\,$ we have the sequence of numbers
\begin{eqnarray}
J_{n,\vec{r}}\,(t) = \left\{ \begin{array}
{l@{\qquad \mbox{if} \qquad}l}
0 & t < |\,\vec{r}\,| - 2\,R_n \;,\\
\mbox{unknown} & |\,\vec{r}\,| - 2\,R_n < t < |\,\vec{r}\,| 
+ 2\,R_n \; ,\\
Q_n(t - \epsilon_n)/\epsilon_n & |\,\vec{r}\,| + 2\,R_n < t\;,
\end{array} \right.
\end{eqnarray}
where $\epsilon_n = |\,\vec{x}_{0,n} - \vec{y}_{0,n}\,|\,$. 
Since $R_n \to 0$ when $n \to \infty\,$, the mid subinterval 
shrinks to the point $t = |\,\vec{r}\,|$, then it is irrelevant 
for our purpose the knowledge of $J_{n,\vec{r}}\,(t)$ in that region. 
Furthermore, since $\epsilon_n \to |\,\vec{r}\,|$ we 
have
\begin{equation}   \label{limJnr}
\lim_{n\to\infty}\, J_{n,\vec{r}}\,(t) =\, \frac{\theta(t-|\,
         \vec{r}\,|)\; Q(t-|\,\vec{r}\,|)}{|\,\vec{r}\,|} \; .
\end{equation}

The limit of $I_n(t)$ can be obtained as follows. For a given 
f\/ixed value $t>0\,$, there exists a natural number $n_0$ 
such that for all $n>n_0\,$ we have $\,t>2\,R_n$. Thus by the 
mean value theorem there also exists a pair of points 
$\vec{x}_{0,n}\,,\, \vec{y}_{0,n} \in \mbox{supp}\,(\rho_n)$ 
such that
\begin{equation}
I_n(t) = \,\frac{Q_n(t) - Q_n(t - \epsilon_n)}{\epsilon_n} \;,
\end{equation}
with $\epsilon_n = |\,\vec{x}_{0,n} - \vec{y}_{0,n}\,| \to 0$ 
when $n \to \infty\,$. Since $Q_n(t)$ has continuous f\/irst 
derivative we can expand it about the f\/ixed value
$t$ up to f\/irst order to obtain
\begin{equation}
I_n\,(t) = \dot{Q}_n\,(t) + O \, (\epsilon_n) \; .
\end{equation}
Therefore the following limit:
\begin{equation}   \label{limIn}
\lim_{n\to\infty} = I_n(t) = \dot{Q}(t)
\end{equation}
holds for all $t>0$.

Finally, using equations~(\ref{A1}), (\ref{limJnr}) 
and (\ref{limIn}) one obtains the radiation reaction
equation~(\ref{rreq2}) for the pointlike limit
case $\,\rho_n(\vec{x}) \to \delta^3(\vec{x})\,$.

%%%%%%%%

\begin{figure} 
\caption{
Two-dimensional schematic f\/igure of a pointlike harmonic oscillator 
at a point $q \in {\cal M}_3$ and its equivalent system on the covering
manifold ${\cal R}^3$, which is formed by an inf\/inite set of identical 
oscillators each one located at a point of the discrete orbit 
$\pi^{-1}\,(q) = {\cal O}_q \subset {\cal R}^3$.
All oscillators interact identically with the scalar f\/ield $\varphi$
and are subject to the same set of initial conditions.}    
\label{capfig1}
\end{figure}

\begin{figure}
\caption{Schematic representation of the curves corresponding 
to the left- and right-hand side of equation~(3.22), whose 
intersection in just one point evinces that this equation has 
a single real positive solution  $\sigma = b $.}
\label{capfig2}
\end{figure}

\begin{figure}
\caption{Graph of the amplitude $Q(t)$ for the overdamped 
harmonic oscillator with $\Gamma=6$ and $\Omega=5$ in a f\/lat 
space-time whose  $t=const$ spacelike section ${\cal M}_3$ is 
endowed with the topology ${\cal R}^2 \times {\cal S}^1$, 
and where we have taken $a=1$ in eq.~(4.1), and as initial
conditions $Q(0)=1\,$ and $\dot{Q}(0)=0$. There are a few 
relative maxima and minima followed by a growth of $Q(t)$. 
This time evolution sharply contrasts with the behavior of the 
amplitude $Q(t)$ for this overdamped oscillator when
${\cal M}_3 = {\cal R}^3$, where there are neither relative 
extrema nor growth of $Q(t)$.
This f\/igure also contains the graph of the function 
$\exp(-bt)\,Q(t)\,$, which within the accuracy of the plot 
conf\/irms an exponential form for the divergent behavior 
suggested by the graph of $Q(t)$. To have this last graph we
have used $b=0.35$, which is the approximate root of 
eq.~(4.2).}
\label{capfig3}
\end{figure}

\begin{figure}
\caption{Behavior of the energy $E(t)$ for the overdamped 
harmonic oscillator with $\Gamma=6$ and $\Omega=5$ in a f\/lat 
space-time whose  $t=const$ spacelike sections ${\cal M}_3$ have 
the topology ${\cal R}^2 \times {\cal S}^1$, and where we have 
taken $a=1$ in eq.~(4.1), and as initial conditions $Q(0)=1\,$ 
and $\dot{Q}(0)=0$. 
The graph for $E(t)$ exhibits a few relative minima and maxima 
followed by an asymptotic divergent behavior. This divergent 
conduct is clearly conform to the behavior of $Q(t)$ when 
$t \rightarrow~\infty\,$.
The time evolution of the energy for the overdamped harmonic 
oscillator in this multiply-connected manifold is in striking 
contrast with the behavior of the energy $E(t)$ in the simply-% 
connected case ${\cal R}^3$, where there is an exponential 
damping of $E(t)$ with no relative extrema. This f\/igure also
contains the graph of the derivative $\dot{E}(t)$, exhibiting 
f\/inite discontinuities at each $t=t_k=k\,a$ ($k$ stands for a 
positive integer), which again do not occur in the simply-connected 
case ${\cal R}^3$. The graph of $\dot{E}(t)$ also reveals that 
$E(t)$ has a f\/inite number of extrema.
The relative minima and maxima of the energy $E(t)$,
the discontinuities of the derivative $\dot{E}(t)$, and the 
asymptotic divergent behavior, which take place in 
multiply-connected f\/lat manifolds, all have a topological 
origin.}
\label{capfig4}
\end{figure}


\begin{references}

\bibitem[*]{internet1} {\sc e-mail address}: german@cbpf.br

\bibitem[\S]{internet2} {\sc e-mail address}: reboucas@cbpf.br

\bibitem[\dag]{internet3} {\sc e-mail address}: teixeira@cat.cbpf.br

\bibitem[\ddag]{internet4} On leave of absence from
Facultad de Ciencias, \ Universidad Nacional de Ingenier\'{\i}a, \ 
Apartado 31 - 139, \  Lima 31 -- Peru.
{\sc e-mail address}: bernui@fc-uni.edu.pe 

\bibitem{BernuiGomeroReboucasTeixeira98} A. Bernui, G.I. Gomero,
M.J. Rebou\c{c}as \&  A.F.F. Teixeira, {\em Phys. Rev. D15\/} 
{\bf 57}, 4699 (1998).

\bibitem{Wolf67} J.A. Wolf, {\em Spaces of Constant Curvature\/},
McGraw-Hill, New York (1967).

\bibitem{Ellis71} G.F.R. Ellis, {\em Gen.\ Rel.\ Grav.\/} {\bf 2},
7 (1971).

\bibitem{EllisSchreiber86} G.F.R. Ellis \& G. Schreiber, 
{\em Phys.\ Lett. A\/} {\bf 115}, 97 (1986).

\bibitem{Gomero97} G.I. Gomero, {\em Fundamental Polyhedron
and Glueing Data for the Sixth Euclidean Compact Orientable 
3-manifold}, CBPF-NF-049/97, Centro Brasileiro de Pesquisas 
F\'{\i}sicas report (1997).

\bibitem{SchwalbThirring64} F. Schwalb \& W. Thirring, {\em Ergeb.\ 
exakten Naturwiss.\/} {\bf 36}, 219 (1964).

\bibitem{Burke71} W. Burke, {\em J. Math.\ Phys.\/} {\bf 12}, 401 
(1971).

\bibitem{AichelburgBeig76} P. Aichelburg \& R. Beig, {\em Ann.\ 
Phys.\ (N.Y.)\/} {\bf 98}, 264 (1976).

\bibitem{AichelburgBeig77} P. Aichelburg \& R. Beig, {\em Phys.\ 
Rev.\ D\/} {\bf 15}, 389 (1977).

\bibitem{Beig78} R. Beig, {\em J.\ Math.\ Phys.\/} {\bf 19}, 
1104 (1978).

\bibitem{Unruh83} W. Unruh, in {\em Gravitational Radiation\/}, Les 
Houches 1982, eds.\ N. DeRuelle \& T.~Piran, North-Holland, 
Amsterdam (1983). 

\bibitem{Stewar83} J. Stewart, {\em Gen.\ Rel.\ Grav.\/} {\bf 15},
425 (1983).

\bibitem{Anderson84} J.L. Anderson,  {\em Gen.\ Rel.\ Grav.\/} {\bf 16},
595 (1984). 

\bibitem{HoenselaersSchmidty89} C. Hoenselaers \& B. Schmidt, 
{\em Class.\ Quantum Grav.\/} {\bf 6}, 867 (1989).

\bibitem{Bernui91} A. Bernui, {\em Appl.\ Anal.\/} {\bf 42},
157 (1991).

\bibitem{Bernui94} A. Bernui, {\em Ann.\ Physik\/} {\bf 3},
408 (1994).

\bibitem{ReboucasTavakolTeixeira98} M.J. Rebou\c{c}as, R.K. Tavakol 
\& A.F.F. Teixeira, {\em Gen.\ Rel.\ Grav.\/} {\bf 30}, 535 (1998).

\bibitem{LachiezeReyLuminet95} M. Lachi\`eze-Rey \& J.-P. Luminet, 
{\em Phys.\ Rep.\/} {\bf 254}, 135 (1995). See also references 
therein.

\bibitem{Kampen51} N.G. van Kampen, {\em Dan.\ Mat.\ Fys.\ Medd.\/} 
{\bf 26}, 16 (1951).

\bibitem{Kramers56} H.A. Kramers, {\em Collected Scientific Papers\/},
p. 845, North-Holland, Amsterdan (1956).

\bibitem{Schutz84} B.F. Schutz, in {\em Relativistic Astrophysics and 
Cosmology\/}, p. 35--97, eds. X. Fustero \& E. Verdaguer, World Scientif\/ic, 
Singapore (1984).

\bibitem{Balbinski85} E. Balbiniski, S.L. Detweiler \& B.F. Schutz 
{\em Mon.\ Not.\ Roy.\ Astron.\ Soc.\/} {\bf 213}, 553 (1985).

\bibitem{Kokkotas86} K.D. Kokkotas \&  B.F. Schutz, {\em Gen.\ Rel.\
Grav.\/} {\bf 18}, 913 (1986).

\bibitem{Richtmyer78} R.D. Richtmyer, {\em Principles of Advanced 
Mathematical Physics. Vol.I\/}, Springer-Verlag, New York (1978).

\bibitem{Maple} K.M. Heal, M.L. Hansen \& K.M. Rickard,  
{\em Maple V Learning Guide\/}, Springer-Verlag, New York (1996).

\bibitem{Heck} A. Heck, {\em Introduction to Maple\/}, 
Springer-Verlag, New York (1993).

\end{references}
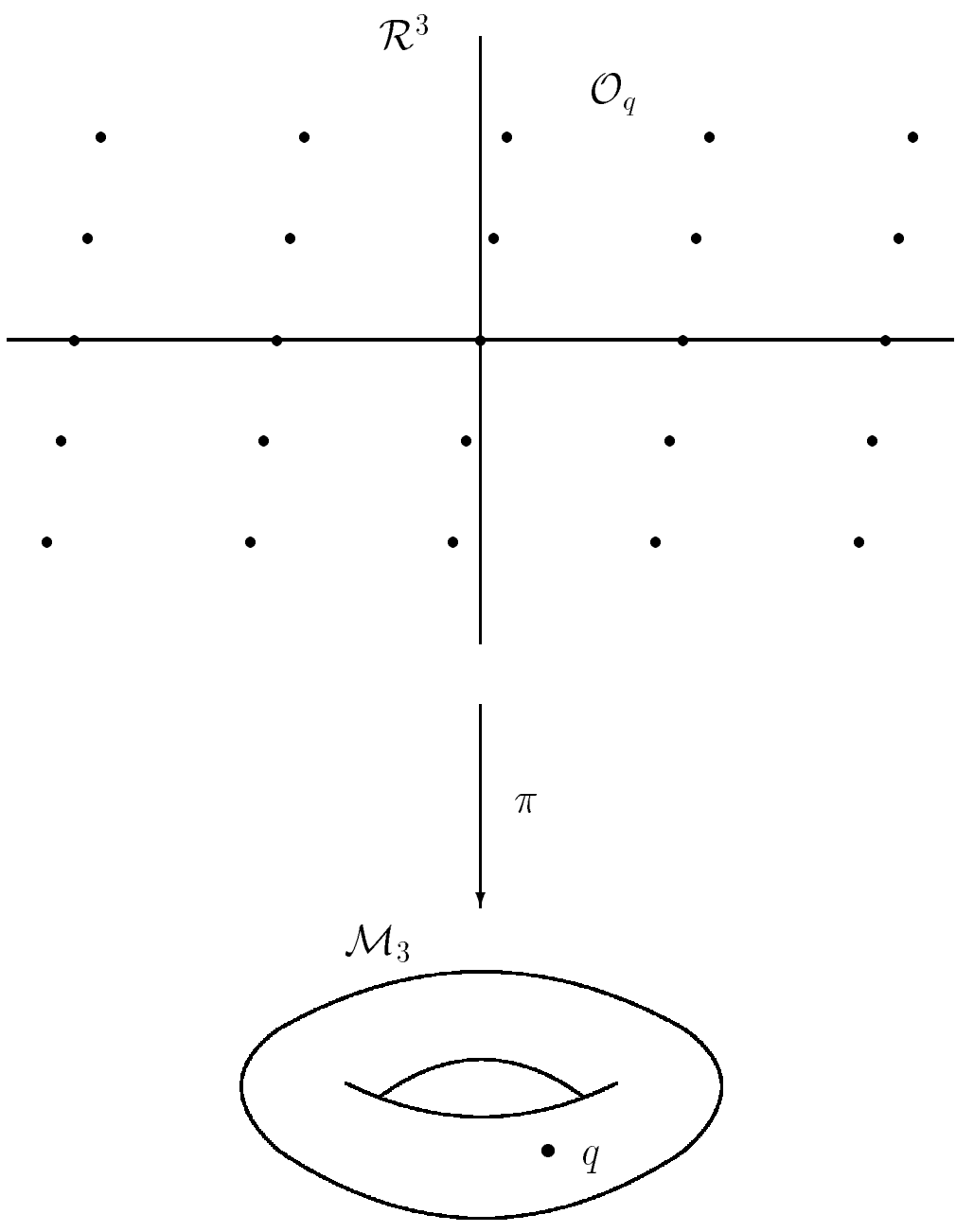
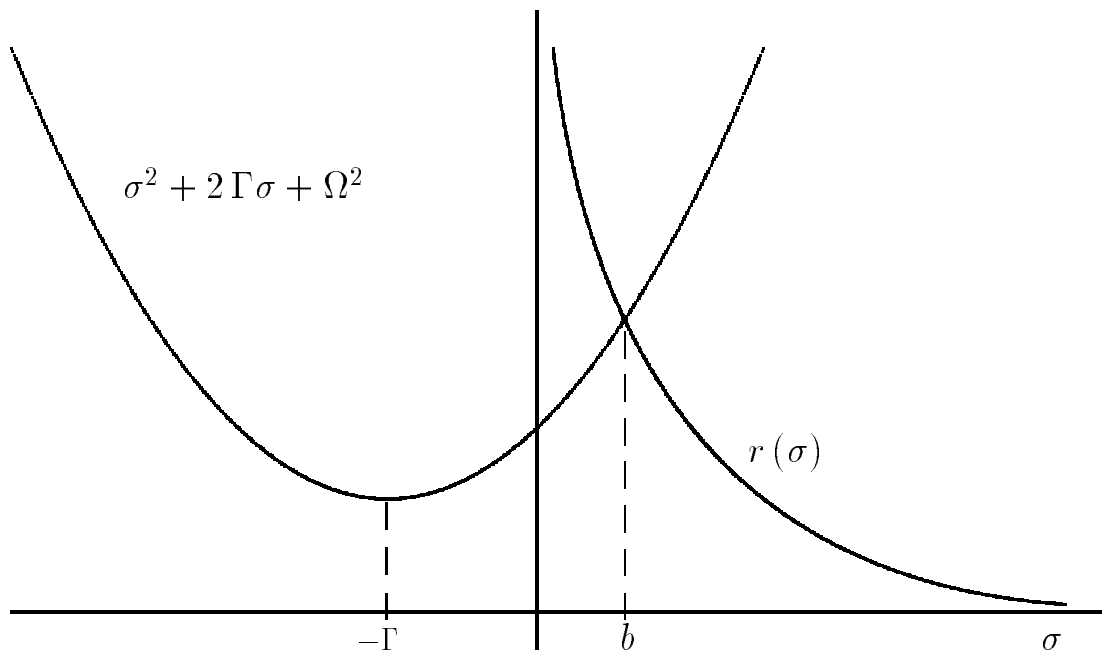
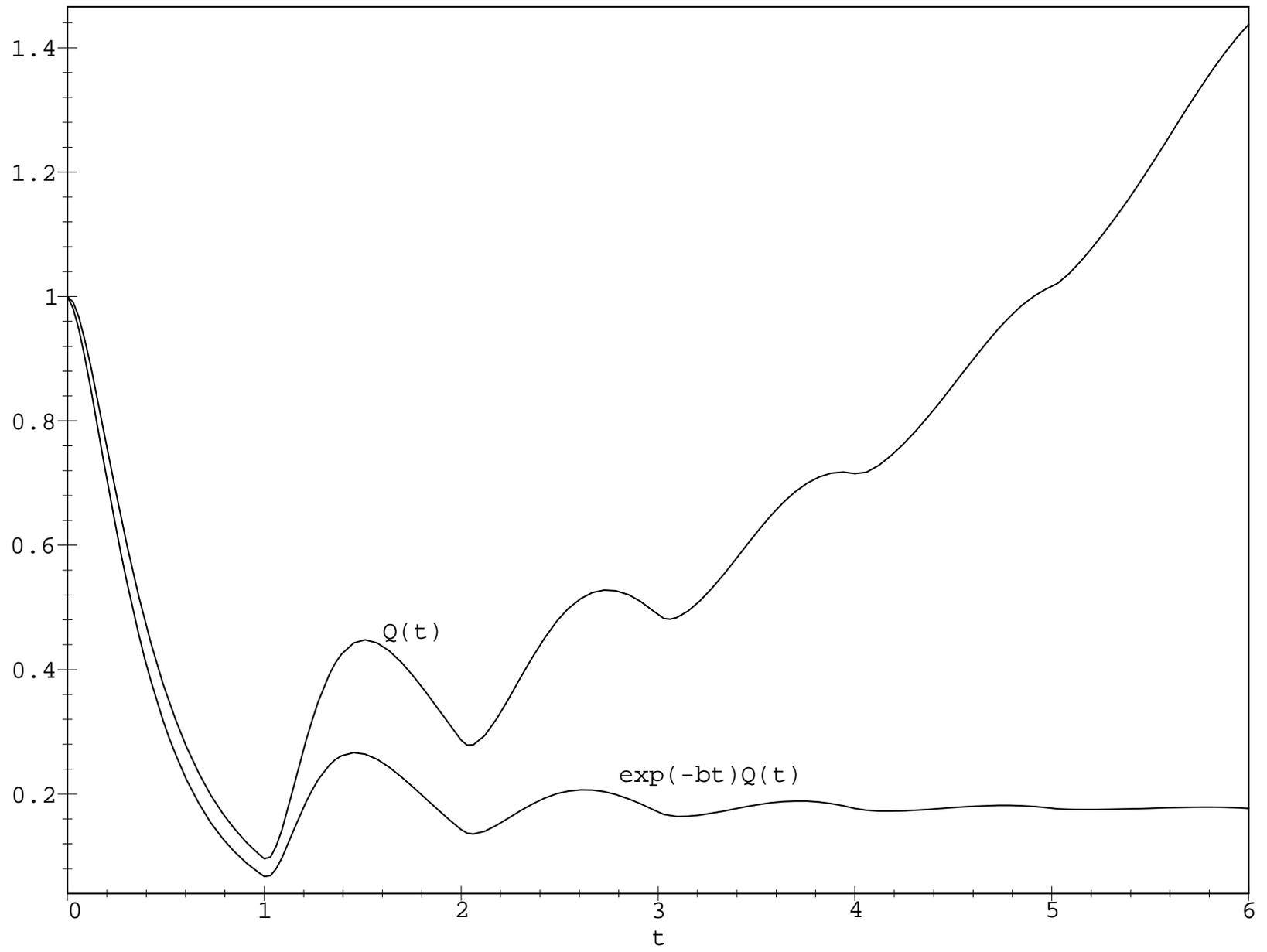
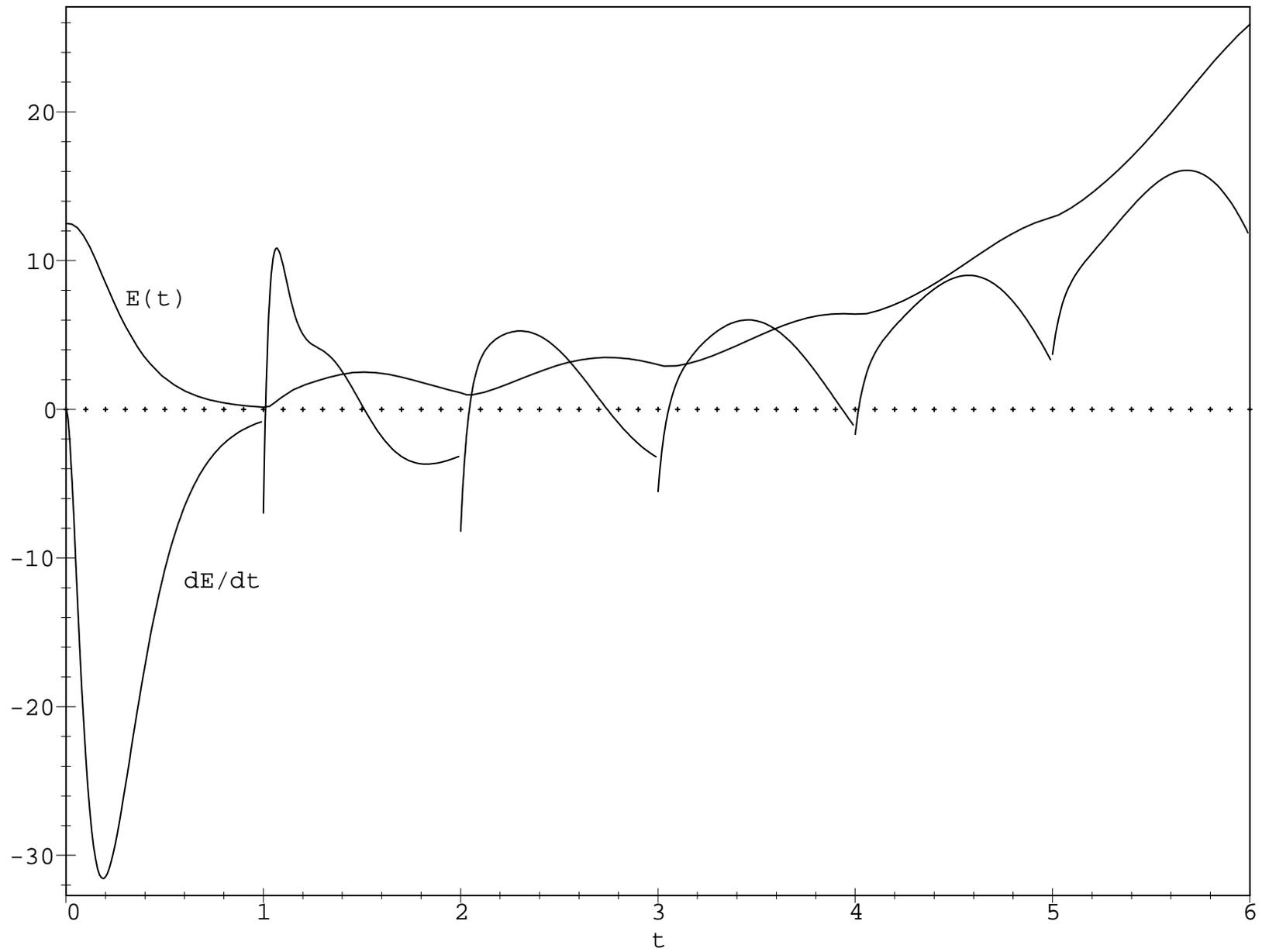
\end{document}